\documentclass{emulateapj}
\usepackage{apjfonts}

\shorttitle{GOODS 850-5}
\shortauthors{Wang, Barger, \& Cowie}

\begin{document}

\title{Ultradeep Near-Infrared Observations of GOODS 850-5\footnotemark[1,2]}

\author{Wei-Hao Wang,\altaffilmark{3,4}
Amy J. Barger\altaffilmark{5,6,7},
and Lennox L. Cowie\altaffilmark{7}}

\footnotetext[1]{Based in part on observations made with the NASA/ESA Hubble Space 
Telescope, obtained at the Space Telescope Science Institute, which is operated 
by the Association of Universities for Research in Astronomy, Inc., under NASA 
contract NAS 5-26555. These observations are associated with program GO 11191.}  
\footnotetext[2]{Based in part on data collected at Subaru Telescope, which is 
operated by the National Astronomical Observatory of Japan.}
\altaffiltext{3}{Jansky Fellow}
\altaffiltext{4}{National Radio Astronomy Observatory,
1003 Lopezville Road, Socorro, NM 87801.  The NRAO is a facility of the National Science
Foundation operated under cooperative agreement by Associated Universities, Inc.}
\altaffiltext{5}{Department of Astronomy, University of Wisconsin-Madison, 
475 North Charter Street, Madison, WI 53706}
\altaffiltext{6}{Department of Physics and Astronomy, 
University of Hawaii, 2505 Correa Road, Honolulu, HI 96822}
\altaffiltext{7}{Institute for Astronomy, University of Hawaii, 
2680 Woodlawn Drive, Honolulu, HI 96822}

\begin{abstract}
GOODS 850-5 is a hyperluminous
radio-faint submillimeter source in the GOODS-N.  Although it is generally 
agreed that GOODS 850-5 is at a high redshift $z\gtrsim4$, its exact 
redshift is unknown.  While its stellar SED 
suggests $z\sim6$, its radio/FIR SED suggests a lower redshift of $z\sim4$.
To better constrain its stellar SED and redshift, 
we carried out \emph{nano-Jansky} sensitivity ultradeep NIR observations between 
1.2 and 2.1 $\mu$m with the \emph{HST} and the 8 m Subaru Telescope.  
Even with such great depths we did not detect GOODS 850-5,
and the results show that it is an extremely curious source.
Between the $K_s$ and 3.6 $\mu$m bands its spectral slope is 
$>3\times$ that of an ERO, and the flux ratio between the two bands is 
$>8\times$ that of Lyman breaks. It is quite challenging to explain this unusually 
red color without a Lyman break (which would imply $z>17$).
It requires a large amount ($M_{\star}\sim10^{11.5} M_{\sun}$) of reddened 
old stars at $z\sim6$, coexisting with an even more extinguished violent 
$\sim2400$--4400 $M_{\sun}$ yr$^{-1}$ starburst,
which does not have any associated detectable rest-frame UV radiation.
We discuss the discrepancy between the NIR and radio/FIR photometric redshifts.  
We conclude that GOODS 850-5 is at 
least at $z>4$ and is more likely at $z\gtrsim6$.  We describe the
unusual properties of GOODS 850-5, including its SED and formation history, 
and we discuss the implications of such massive $z>6$ galaxies.
\end{abstract}


\section{Introduction}
The Rayleigh-Jeans portion of the dust spectral energy distribution (SED) of 
IR-luminous galaxies produces a strong negative $K$-correction
and makes the observed submillimeter flux of such galaxies
almost invariant at $z>1$ to $z\sim10$ \citep{blain93}.
Thus if there are IR-luminous galaxies at high redshift, observations at
submillimeter wavelengths are a powerful way to find them.
However, to date, all but one of the identified submillimeter galaxies (SMGs), 
other than those around luminous quasars, are at redshifts lower than 4,
likely because of the limited resolution of the
current submillimeter instruments and the limited sensitivity 
of the current radio instruments, which are needed to locate the sources.

The Submillimeter Common-User Bolometer Array (SCUBA) on the 
single-dish James Clerk Maxwell Telescope (JCMT) resolved 20\%--30\% 
of the submillimeter extragalactic background light (EBL) into point 
sources brighter than $\sim2$ mJy at 850 $\mu$m 
(\citealp{smail97,barger98,hughes98,barger99,eales99}).  Because of the low resolution
of JCMT ($\sim15\arcsec$ at 850 $\mu$m), identifications of the submillimeter
sources have to assume the radio--far-infrared (FIR) correlation in local galaxies 
\citep[see, e.g.,][]{condon92} and rely on radio interferometry
to pinpoint the location of the submillimeter emission. 
Optical spectroscopy of radio identified SMGs shows that they are 
ultraluminous sources ($>10^{12}$ $L_{\sun}$, corresponding to a star formation rate
of $10^2$--$10^3$ $M_{\sun}$ yr$^{-1}$) peaking at $z\sim2$--3 and that 
they dominate the total star formation in this redshift range \citep{chapman03,chapman05}.  
However, the positive $K$-correction of the radio synchrotron emission 
makes the radio wavelength insensitive to high-redshift galaxies, and radio 
observations can only identify 60\%--70\% of the blank-field
submillimeter sources \citep{barger00,ivison02}.  
The radio unidentified SMGs are commonly thought 
to be at redshifts higher than the radio detection limit 
(typically $z\gtrsim3$--4) but there has been a lack of direct 
evidence for such a high-redshift radio-faint submillimeter population.
To date, there is only one spectroscopically confirmed SMG at $z>4$
\citep[$z=4.547$]{capak08}.  Luminous radio-faint SMGs contribute $\sim10\%$ to 
the submillimeter EBL measured by the \emph{COBE} satellite \citep{puget96,fixsen98}.
If most of these galaxies are indeed at high redshifts, then this implies a
large amount of star formation in massive high-redshift systems.

\begin{figure*}[ht!]
\epsscale{1.16}
\plotone{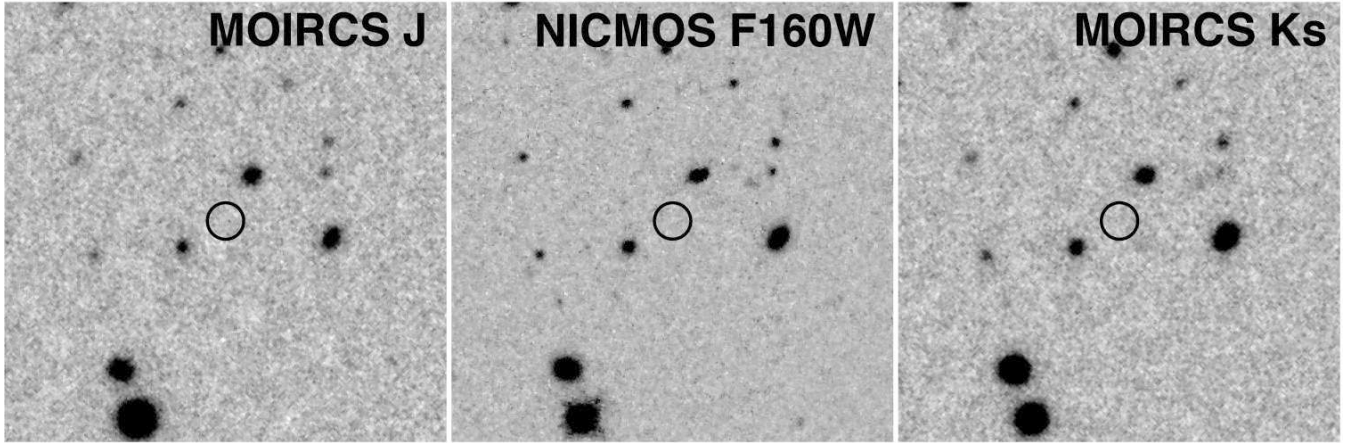}
\caption{Our new ultradeep 
NIR $J$, F160W, and $K_s$ band images of GOODS 850-5.
Each panel has a $24\arcsec$ width.  North is up.  
The three images have identical surface brightness scales (in $f_\nu$).
The $2\arcsec$ diameter circles 
mark the SMA position from W07, which has an uncertainty of $0\farcs2$.
\label{pic_nicmos_moircs}}  
\end{figure*}

With recent developments in submillimeter interferometry,
it is now possible to directly locate submillimeter sources
without relying on radio interferometers.  \citet{younger07} 
observed a sample of SMGs with the Submillimeter Array (SMA).
Several of their SMA detections are radio faint and consistent with
being at redshifts higher than the radio identified SMGS.
We have also been carrying out a program specifically targeting radio-faint submillimeter 
sources with the SMA.  In \citet[][hereafter W07]{wang07} we reported our 
first identification in this program, GOODS 850-5. 
GOODS 850-5 was detected in our JCMT
SCUBA jiggle-map survey of the Great Observatories Origins 
Deep Survey-North (GOODS-N, \citealp{giavalisco04a}) with an 
850 $\mu$m flux of $12.9\pm2.1$ mJy \citep{wang04}.
It was also detected in the combined jiggle and scan map
of the GOODS-N (GN 10, see \citealp{pope06} and references therein).  
It is the second brightest submillimeter 
source in our jiggle-map catalog of the GOODS-N and has a total IR
luminosity of $\sim2\times10^{13}$ $L_{\sun}$.  It did not have a 
$5 \sigma$ radio counterpart in the deep Very Large Array (VLA) 
1.4 GHz catalogs of \citet{richards00} and \citet{biggs06}.  

In W07, an SMA 880 $\mu$m detection of GOODS 850-5 was obtained
and its counterpart was found to be extremely faint in the optical and 
near-IR (NIR).  \citet[][hereafter D08]{dannerbauer08} soon followed up with new 1.25 mm 
and 20 cm detections of GOODS 850-5.  It turns out that the correct \emph{Spitzer} 
counterpart had already been suggested by \citet{pope06} but this identification
was not confirmed and the high redshift of the source was not realized 
until the accurate position was obtained by W07.
However, the exact redshift of GOODS 850-5 is
unclear.  Both W07 and D08 found that the NIR SED is consistent
with a galaxy at $z\sim6$, and the submillimeter and radio SED is consistent with
$z\sim4$.  While both groups agree that this is a high-redshift SMG, 
W07 favors $z\sim6$ but D08 favors $z\sim4$ based on the radio--FIR correlation.  
The $z\sim6$ redshift suggested
by W07 is based on a featureless power-law continuum in the IRAC bands
and a non-detection in a relatively shallow $K_s$ band image.  However,
it is fair to say that 
this photometric redshift is not a secure one, which would require the detection of
at least one prominent spectral feature.  To obtain a better constraint on the
redshift, we carried out ultradeep NIR imaging in the $J$, $H$, and $K_s$ bands, 
hoping to detect the stellar continuum at $\lesssim2$ $\mu$m.  
However, we found instead that the SED of GOODS 850-5 at 1--3 $\mu$m 
is quite weird---it is not detected even at a 5 nJy sensitivity at 1.6 $\mu$m
(compared with its $\sim1$ $\mu$Jy flux at 3.6 $\mu$m).  Explaining this is challenging, 
but our analyses show that it greatly strengthens the previous W07 suggestion of $z\gtrsim6$.

In this paper we present our new ultradeep NIR imaging observations of 
GOODS 850-5 and complete analyses of the likelihood of its redshift.
The new observations and data used in this paper are described in \S~\ref{sec_obs}.
Photometric redshift analyses in the optical, NIR, FIR, and radio are described in \S~\ref{sec_photoz}.
The mass and age of the stellar population and the star formation rate of GOODS 850-5
are estimated, and the nature of this galaxy is discussed in \S~\ref{sec_property}.  
The implications for galaxy evolution are
discussed in \S~\ref{sec_discussion}.  We summarize and make some final
remarks in \S~\ref{sec_summary}.
The cosmological parameters adopted in this paper are $H_0=71$ km s$^{-1}$ Mpc$^{-1}$,
$\Omega_{\Lambda}=0.73$, and $\Omega_M=0.27$.

\begin{figure*}
\epsscale{1.17}
\plotone{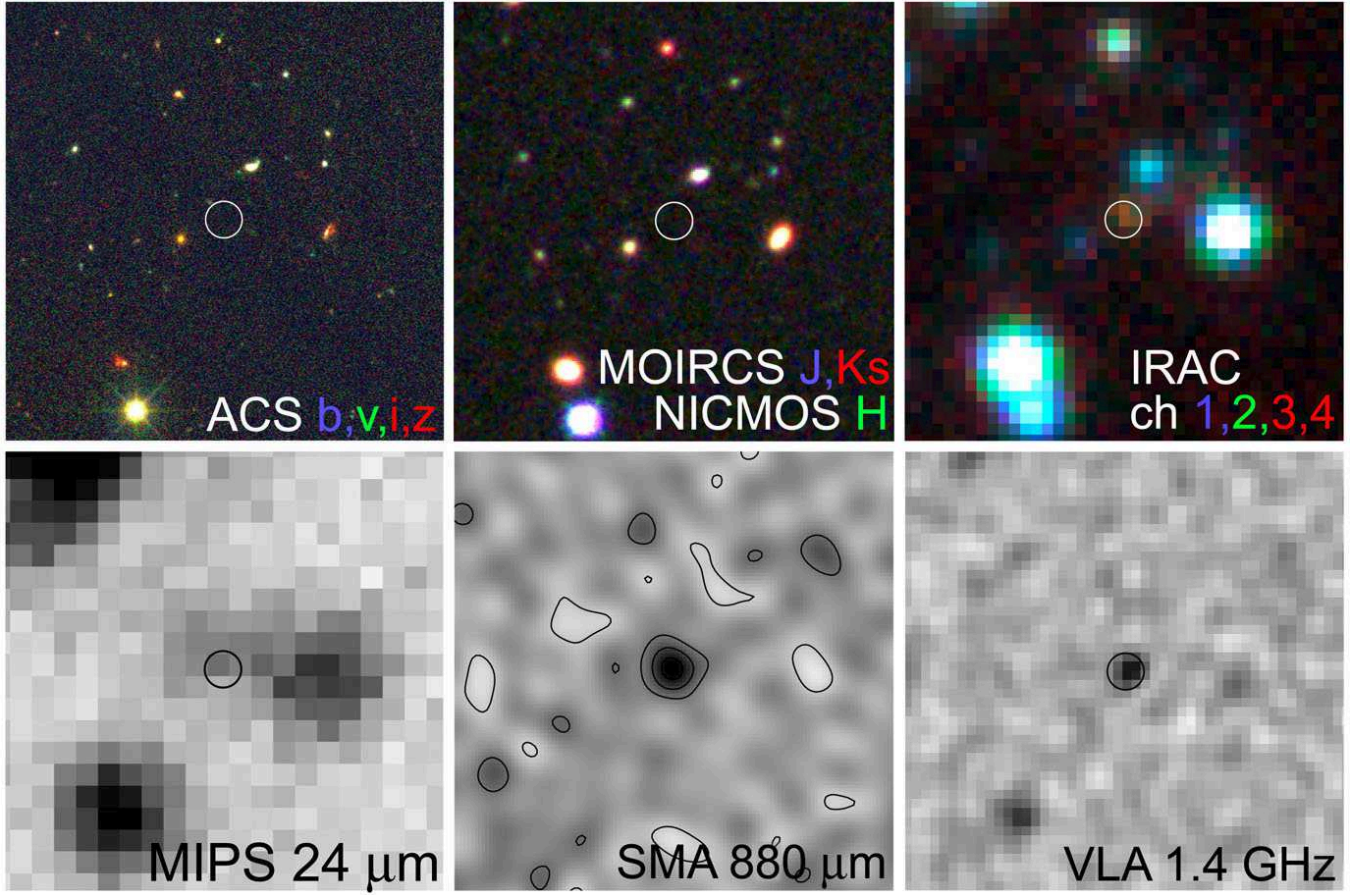}
\caption{Multiwavelength images of GOODS 850-5.  
Each panel has a $24\arcsec$ width.  North is up.  
The SMA position is labeled with $2\arcsec$ diameter circles and has an
uncertainty of $0\farcs2$.
The color codes in the optical, NIR, and IRAC images are
labeled in the pictures.  Grayscale images have negative color scales.
The MOIRCS and NICMOS images are from
this work and the VLA image is from D08 and G.\ Morrison (in preparation).  
The rest are from W07.
\label{thumbnails}}  
\end{figure*}

\section{New NIR Imaging and Existing Data}\label{sec_obs}

\subsection{\emph{HST} NICMOS F160W Imaging}
We observed GOODS 850-5 with the Near Infrared Camera and Multi-Object 
Spectrometer (NICMOS) on the \emph{Hubble Space Telescope} (\emph{HST})
in Cycle 16.  We used the NIC3 camera and the F160W filter to obtain the highest
sensitivity.  The observations were made in November and December 2007, 
containing a total of 16 \emph{HST} orbits in four visits.  In each orbit, we made only one 
exposure that is as long as possible ($\sim47$ minutes) in order to minimize the 
contribution of read noise.  The exposures were dithered to provide 
0.5 pixel sampling.

The data reduction was carried out in the Interactive Data Language (IDL) 
environment.  We started the reduction with the standard pipeline-calibrated images 
provided by the \emph{HST} Archive, which are flux calibrated and have instrumental
singatures removed.  We applied a background subtraction to each exposure
by fitting the object masked image with a smooth polynomial surface.  The brightest
cosmic ray hits and hot pixels were removed in each image with a spatial sigma
filtering.  The offsets between images were determined by measuring the positions
of the detected objects in the central part of the field of view.
We did not attempt to correct for the optical distortion of NIC3.
The 16 images were then drizzled to a common grid of $0\farcs1$ per pixel
(two times finer than the original pixel of NIC3) and combined to form a deep image.
The absolute astrometry was matched to the GOODS-N ACS catalog \citep{giavalisco04a} 
but corrected for the $0\farcs4$ offset in the ACS frame to match the radio frame
\citep{richards00}.
Before the images were combined, another sigma filtering was applied to drizzled 
pixels in the same coordinate grids to remove fainter cosmic ray hits.  
When the images were combined, each pixel was weighted by its exposure time
and each image was inverse-weighted by the background brightness to achieve
the highest S/N.  The reduced image is shown in Figure~\ref{pic_nicmos_moircs}.
The final image has an extremely deep 1 $\sigma$ sensitivity of 4.9 nJy.  
Surprisingly, even with such a great depth, we are not able to detect 
GOODS 850-5. 

\subsection{Subaru MOIRCS $J$ and $K_s$ Imaging}

We obtained extremely deep $J$ and $K_s$ images of GOODS 850-5
with the Multi-Object InfraRed Camera and Spectrograph \citep[MOIRCS, ][]{ichikawa06}
on the 8 m Subaru Telescope on Mauna Kea.  MOIRCS contains two 2k~$\times~$2k 
HAWAII2 detectors, covering a field of view of $\sim 4\arcmin \times 7\arcmin$ with a 
pixel scale of $0\farcs117$.  The images used here were taken in two large MOIRCS 
imaging campaigns in the GOODS-N led by Japanese groups and our group based in
Hawaii between December 2005 and January 2008.  All the $J$ band exposures and 
approximately half of the
$K_s$ band exposures were made by Japanese investigators and were obtained 
from the Subaru public archive.  Parts of the Japanese data were published in
\citet{kajisawa06}.  The rest of the $K_s$ exposures were made by our group and will
be published elsewhere.  The majority of the observations were performed under photometric 
conditions with excellent seeing 
of $0\farcs25-0\farcs6$.  A very small fraction of the data has a large extinction 
of $>0.5$ magnitude or poor seeing of $>0\farcs7$, and they were excluded in 
this work.

The MOIRCS images were reduced with the IDL based 
SIMPLE Imaging and Mosaicking Pipeline (SIMPLE, W.-H.\ Wang 2008,
in preparation\footnotemark[8]).
Images within a dither set (typically 20--30 minutes in length) were flattened
using an iterative median sky flat in which a simple median sky was first
derived to flat the images and then a second median sky was derived by masking
all the detected objects using the flattened images.  After the images were flattened,
the residual sky background was subtracted with a 
smooth polynomial surface.  The brightest cosmic ray 
hits were removed by a spatial sigma filtering in each flattened image.
MOIRCS produces almost nearly circular fringes in
roughly half of the images.  The fringes were modeled in polar coordinates 
where they are nearly perfect straight lines and were subtracted from the images in
the original Cartesian coordinates.  

\footnotetext[8]{also see http://www.aoc.nrao.edu/$\sim$whwang/idl/SIMPLE/index.htm}

The package SExtractor \citep{bertin96} was used to measure object positions 
and fluxes in each flattened, sky subtracted, and fringe removed image in 
a dither set.  The first-order derivative of the optical distortion function was 
derived by measuring the offsets of each object in the dither sequence as a 
function of location in the images.  Absolute astrometry was obtained by matching 
the detected objects to a reference catalog constructed with brighter and compact
objects in the GOODS-N ACS catalog
(matched to the radio frame) and the GOODS-N SuprimeCam catalog \citep{capak04}.
The images were then warped directly from the raw frames to a 
common tangential sky plane with a sub-pixel accuracy.  
All projected images were weighted by their sky transparencies, exposure 
times, inverse background brightnesses, and pixel-to-pixel efficiencies
(i.e., flat field) to obtain optimal S/N.
The weighted images were then combined to form a large mosaic.  When images 
from a dither set were combined, a sigma filter was applied to pixels that have the 
same sky positions to further remove fainter cosmic rays.  
The images were calibrated by observing various UKIRT Faint Standards \citep{hawarden01} 
at least every half night on each detector. 
Data taken under nonphotometric conditions and poorly calibrated archive data 
were recalibrated with photometric data taken by our group.  

The final mosaics have weighted exposure times (relative to 
median sky brightness and transparency) of 13.2 and 23.7 hours in the
$J$ and $K_s$ bands, respectively, at the location of GOODS 850-5.
The image qualities are very good, with FWHMs of $0\farcs46$ at $J$ and
$0\farcs42$ at $K_s$.  The rms astrometry error between the MOIRCS source positions 
and the ACS/SuprimeCam reference catalog is $0\farcs08$.
The reduced MOIRCS images in the region around GOODS 850-5
are shown in Figure~\ref{pic_nicmos_moircs}.
The $1 \sigma$ sensitivity limits at $J$ and $K_s$ are both 14 nJy. 
As with the F160W image, GOODS 850-5 is not detected in the
$J$ and $K_s$ images.

\subsection{NIR Photometry}
Although GOODS 850-5 is not detected in any of the above ultradeep NIR 
images, it is useful to measure its fluxes at the SMA position (J2000 = 12:36:33.45,
+62:14:08.65, W07) to determine whether there is any low level flux recorded.
It is also important to obtain realistic flux limits/errors in these bands for the use of
photometric redshift fitting.  We placed apertures with diameters that are 1.5 times the 
FWHMs of compact objects in the fields ($0\farcs46$, $0\farcs3$, and $0\farcs42$ in $J$, F160W,
and $K_s$, respectively) at the SMA position to measure the fluxes of GOODS 850-5.
For each image, the background was estimated in a $3\arcsec$ area around GOODS 850-5 after
detected objects and GOODS 850-5 itself were masked.
The measured $J$ and F160W fluxes are negative, and the $K_s$ flux is 0.65 $\sigma$.
To estimate flux uncertainties, we first masked all detected objects in the images
and then randomly placed the apertures in the neighborhood of GOODS 850-5.
The dispersions in the random aperture fluxes are considered as flux uncertainties
in these three bands.  This procedure is necessary especially for the two MOIRCS bands,
where the images were not drizzle combined and therefore the noise is more correlated 
between pixels.  Such flux errors also include the uncertainties in background subtraction
and confusion noise from faint undetected sources in the field.
We summarize the measurements in Table~\ref{tab1}.

\begin{deluxetable}{lc}[hb!]
\tablecaption{NIR Photometry of GOODS 850-5\label{tab1}}
\tablehead{\colhead{Wave Band} & \colhead{Flux (nJy)}}
\startdata
MOIRCS $J$ & $-0.8\pm13.8$ \\
NICMOS F160W & $-2.75\pm4.90$\\
MOIRCS $K_s$ & $9.1\pm13.9$
\enddata
\end{deluxetable}

\subsection{Existing Data}
GOODS 850-5 is not detected in the GOODS-N \emph{HST} ACS images but is clearly 
detected in the \emph{Spitzer} IRAC 3.6--8 $\mu$m and MIPS 24 $\mu$m images
(GOODS \emph{Spitzer} Legacy Program, M.\ Dickinson et al.\ 2008, in preparation).  
Its IRAC and MIPS fluxes were first measured by \citet{pope06}.  Because GOODS 850-5
is blended with brighter nearby IRAC and MIPS sources, in W07 we remeasured the IRAC 
and MIPS fluxes with a PSF fitting method to better isolate its flux.
In this work, we adopt the W07 results.  (The flux errors in W07 and here include the
uncertainties in the deblending processes.)  
GOODS 850-5 is not detected in
the MIPS 70 $\mu$m and 160 $\mu$m bands \citep{frayer06,huynh07}.
We adopt the 1 $\sigma$ limits of 2 mJy at 70 $\mu$m and 30 mJy at 160 $\mu$m.

In the submillimeter, we adopt the SCUBA 850 $\mu$m jiggle map flux of Wang et al.\ (2004)
and the SMA 880 $\mu$m flux of W07.  GOODS 850-5 was detected by D08 with
the IRAM Plateau de Bure Interferometer at 1.25 mm with a flux of $5.0\pm1.0$ mJy.
In W07, we used the older Very Large Array (VLA) 20 cm image of 
\citet[also see \citealt{richards00}]{biggs06}, in which GOODS 850-5 has a flux of
$18.7\pm8.0$ $\mu$Jy\footnotemark[9].  G.\ Morrison et al.\ (in preparation) obtained a deeper 
VLA 20 cm image of GOODS-N.  GOODS 850-5 was detected with a highly significant
flux of $34.4 \pm 4.2$ in the new VLA image (D08).  Here we adopt the latest D08 values at 
1.25 mm and 20 cm.  The photometry used in this work is summarized in Table~\ref{tab2}.  
Figure~\ref{thumbnails} show the multicolor images of GOODS 850-5.  
Comparing to the figure shown in W07, the main difference is the new Subaru 
and \emph{HST} NIR images and the new VLA 20 cm image.

\footnotetext[9]{This flux was measured at the location of the SMA position, which is
$\sim0\farcs3$ away from the best-fit VLA position.  G.\ Morrison (2008, personal 
communication) and R. Ivison (2008, personal communication) found best-fit VLA fluxes 
of $24.1\pm5.8$ and $23.8\pm5.9$ $\mu$Jy, respectively, from the image of \citet{biggs06}.}

\begin{deluxetable}{lcc}
\tablecaption{Photometric Data of GOODS 850-5\label{tab2}}
\tablewidth{220pt}
\tablehead{\colhead{Wave Band} & \colhead{Flux ($\mu$Jy)} & \colhead{Reference}}
\startdata
ACS F435W & 		$-0.013\pm 0.004$ &	2 \\
ACS F606W & 		$-0.004 \pm 0.003$ &	2 \\
ACS F775W & 		$0.001 \pm 0.006$ &	2 \\
ACS F850LP &		$-0.009 \pm 0.009$ &	2 \\
MOIRCS $J$ &        $-0.0008\pm0.014$ & 1\\
NICMOS F160W &   $-0.002.8\pm0.0049$ & 1\\
MOIRCS $K_s$ &    $0.0091\pm0.014$ & 1\\
IRAC 3.6 $\mu$m &	$1.14 \pm 0.14$ &		2 \\
IRAC 4.5 $\mu$m &	$1.64 \pm 0.13$ &		2 \\
IRAC 5.8 $\mu$m &	$2.33 \pm 0.24$ &		2 \\
IRAC 8.0 $\mu$m &	$5.37 \pm 0.37$ &		2 \\
MIPS 24 $\mu$m &	$46.3 \pm 9.2$ &		2 \\
MIPS 70 $\mu$m &	$<2000$ &		3 \\
MIPS 160 $\mu$m &	$<30000$ &		4 \\
SCUBA 850 $\mu$m & $12900 \pm 2100$ &	5 \\
SMA 880 $\mu$m &	$12000 \pm 1400$ &	2 \\
IRAM 1.25 mm &	$5000\pm1000$ &		6 \\
VLA 20 cm &		$34.4\pm4.2$ &		6 
\enddata
\tablerefs{(1) this work; (2) \citet[W07]{wang07}; (3) \citet{frayer06}; (4) \citet{huynh07}; 
(5) \citet{wang04}; (6) \citet[D08]{dannerbauer08}.}
\end{deluxetable}

\section{SED and Redshift of GOODS 850-5}\label{sec_photoz}

\subsection{Non-detections in $J$, F160W, and $K_s$}
The results of our new NIR imaging are surprising and not particularly 
easy to understand.
GOODS 850-5 is clearly detected in the IRAC 3.6 $\mu$m band at 8 $\sigma$.
Despite the great sensitivities at F160W and $K_s$, which are respectively 
29 and 10 times higher than those at 3.6 $\mu$m, GOODS 850-5 is not detected.  
This reveals an extraordinarily red SED in the NIR.  The 1 $\sigma$ $K_s$ flux 
upper limit (23 nJy) implies a $S_{3.6\mu\rm m}/S_{Ks}$ flux ratio of $>50$ or 
a spectral slope of $\alpha < -7.8$.  For comparison, objects with $R-K>5$, 
corresponding to $\alpha < -2.5$, are called ``extremely red objects''
\citep[EROs, e.g.,][]{mccarthy04}.  The spectral slope of GOODS 850-5 is
more than three times that of EROs, and the $K_s-3.6$ $\mu$m color 
(in AB scale) of GOODS 850-5 is $\gtrsim2$ higher than the $I-J$ color 
of Hu-Ridgway 10, which is a prototype dusty starburst ERO \citep{hr10}.
High redshift UV emitting objects are commonly selected with red 
colors of $\gtrsim2$ (in the AB system) between two adjacent optical filter bands
(the ``Lyman-break'' technique, e.g., \citealt{steidel99,fan01}),
corresponding to $\alpha \lesssim -10$.  The flux ratio of GOODS 850-5 
between $K_s$ and 3.6 $\mu$m is $\sim8$ times that of a minimal
Lyman break.
(A Lyman break between $K_s$ and 3.6 $\mu$m would imply $z>17$.)

\begin{figure*}
\epsscale{0.85}
\plottwo{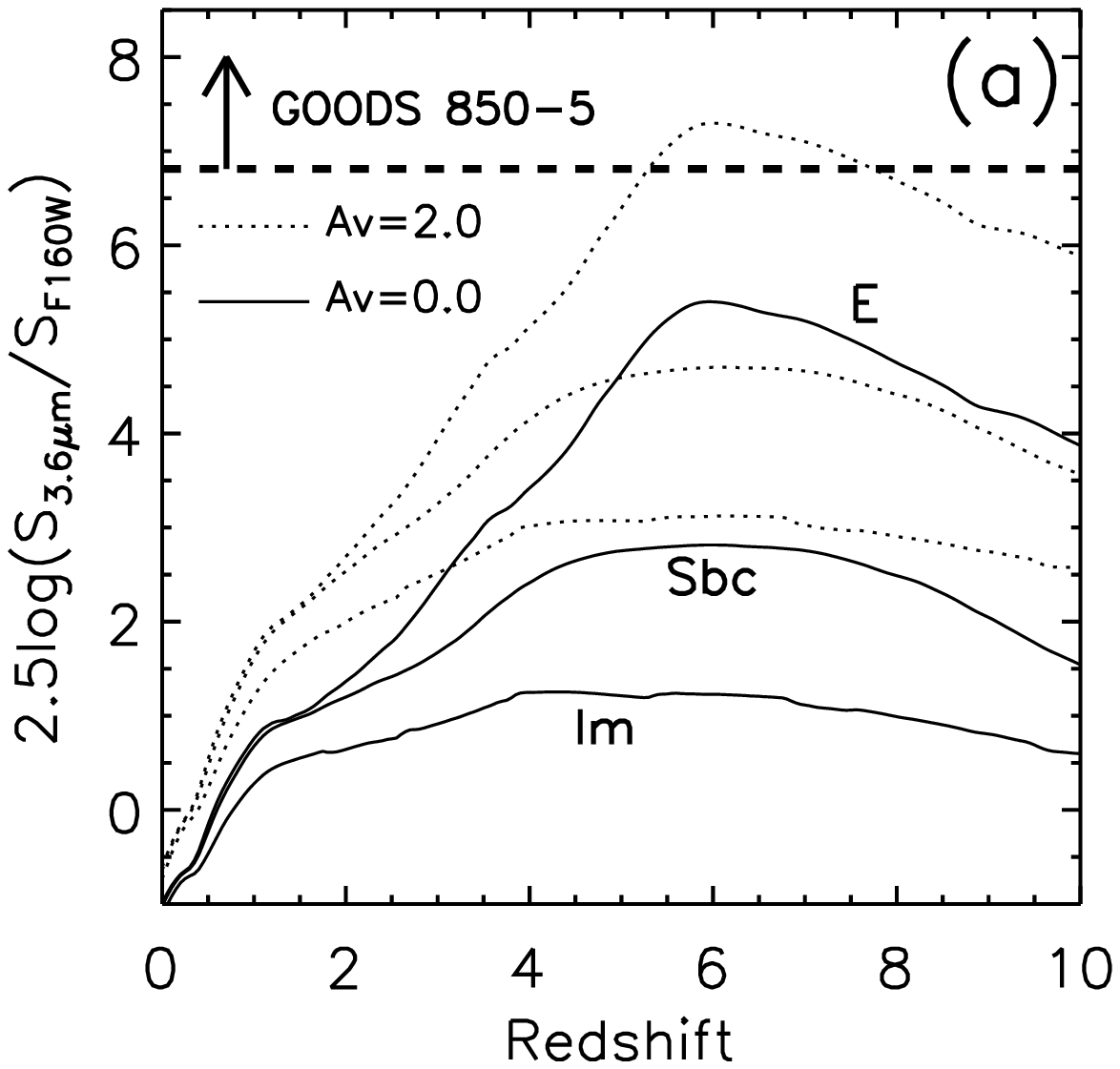}{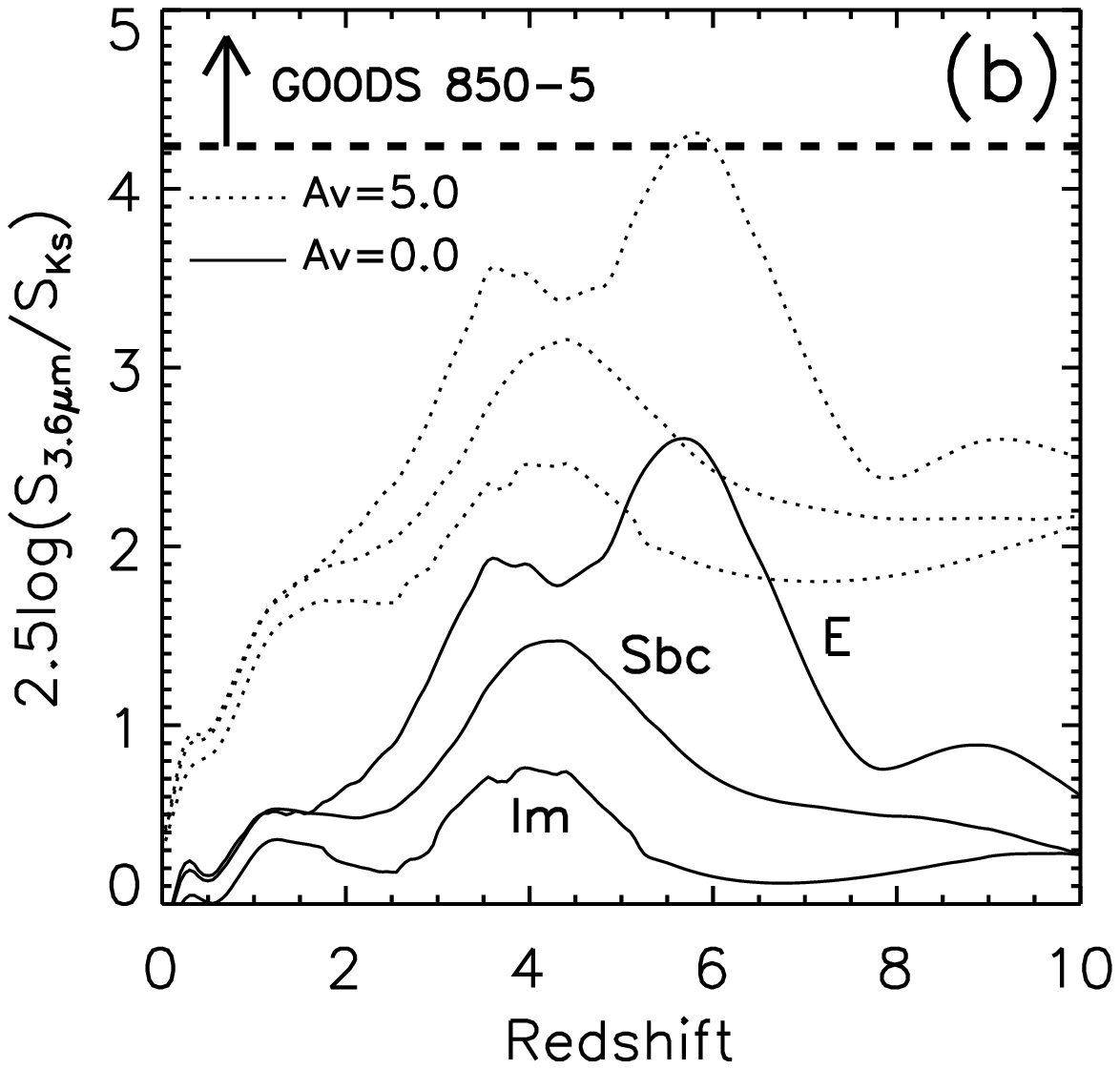}
\caption{NIR colors of GOODS 850-5 and galaxies between $z=0$ and 10: 
(a) colors between the F160W and the 3.6 $\mu$m bands; (b) colors
between the $K_s$ and 3.6 $\mu$m bands.  Solid curves are colors of E, Sbc, and Im
galaxy types in \citet{cww80}, without reddening.  Dotted curves are the 
same types of galaxies with $A_V=2.0$ (a) and $A_V=5.0$ (b).
We adopt the extinction law in \citet{calzetti00}.  Horizontal dashed lines are the
lower limits for the colors of GOODS 850-5, derived from the observed $1\sigma$
upper limits of its F160W and $K_s$ fluxes.
\label{nir_colors}}  
\end{figure*}

The extremely red color of GOODS 850-5 sets a strong constraint on its redshift.
It can be seen in Figure~\ref{nir_colors} that it is very difficult to explain the 
observed colors between 1.6 and 3.6 $\mu$m with galaxies at $z<5$ even with a 
considerable amount of reddening.  Simply based on the two colors
in Figure~\ref{nir_colors}, $z\sim6$ appears to be the most likely 
redshift for GOODS 850-5.  However, we can use all the filter bands simultaneously
to obtain better redshift estimates with the photometric redshift technique.

\subsection{NIR Photometric Redshift}\label{sec_nir_photoz}
In W07 we derived photometric redshifts for GOODS 850-5 and found that
galaxies at $z\sim6$ provided the best fits.  Here we used our new photometric
data in the NIR to improve our photometric redshift estimates.  
We used all the \emph{Spitzer} IRAC, 
\emph{HST} ACS and NICMOS, and Subaru MOIRCS data and the latest version of the 
Hyperz package \citep{bolzonella00}\footnotemark[10] for this calculation.  
In each band, we added a 5\% of zeropoint error quadratically.
We used zeros to replace negative fluxes in the four ACS bands and the $J$ and F160W 
bands, since negative fluxes are not physical.  (The F775W flux is consistent
with zero.)  We used the non-zero $K_s$ flux.
The photometric redshifts were derived independently
with two SED template sets.  The first is the latest stellar population synthesis
model of \citet[BC03]{bc03}.  The second contains the empirically observed galaxy 
spectra of \citet{cww80} from E to Im types and the starburst spectra of \citet{kinney96}.
Both SED sets are widely used by the optical extragalactic community for
photometric redshift estimates.  The combination of the latest Hyperz with the
BC03 model has the nice feature of estimating ages and stellar masses of galaxies.
Hyperz also limits the age of the galaxies not to exceed the age of the universe.

\footnotetext[10]{also see http://www.ast.obs-mip.fr/users/roser/hyperz/}

We adopted the standard extinction law of \citet{calzetti00}. 
There is a subtlety in the maximum extinction to be allowed.  
Explaining the extremely red color of GOODS 850-5 between 
F160W/$K_s$ and 3.6 $\mu$m requires a strong Balmer break 
(rest wavelength 4000 \AA) at a high redshift of $z>5$ (or a Lyman break at $z>16$).   
However, if we allow very large extinctions, this red color might be
reproduced with a highly reddened low-redshift galaxy.
For example, \citet{mobasher05} suggest that the very red $z-J$ color of the object 
HUDF-JD2 may come from a Lyman break at $z\sim6.5$.  
On the other hand, the mid-IR (MIR) observations of 
\citet{chary07} suggest that HUDF-JD2 is at $z\sim2$ and the red $z-J$ color comes
from an extinction of $A_V\sim4.9$ (also see \citealp{fontana06,dunlop07}).  
Spectroscopically confirmed SMGs have
a typical extinction of $A_V\sim1$--3 \citep{smail04,swinbank03}, but the
spectroscopic sample may be biased toward less reddened systems.
Photometric redshift fitting to SMGs without spectroscopy generally gives 
$A_V\sim0$--5.  From these examples, $A_V<6$ seems to be a reasonable limit 
for high-redshift SMGs.  (For heavier extinctions, the shape of the extinction curve becomes a 
much more important uncertainty than the $A_V$ value itself.)
We adopted this limit for our photometric redshift analyses.
Later we will show that, unlike in the case of HUDF-JD2, a $z<3$ redshift with 
a large extinction is ruled out for GOODS 850-5 by its radio and FIR SED.

\begin{figure}
\epsscale{1.1}
\plotone{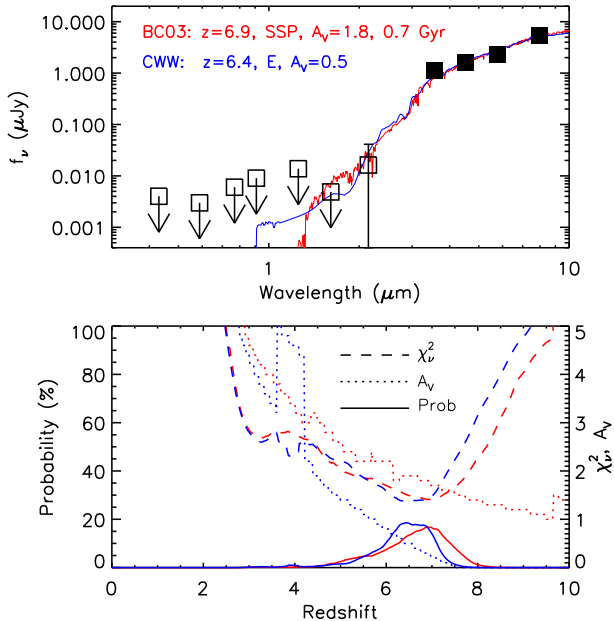}
\caption{NIR photometric redshift results.  The top panel shows the observed
SED between 0.4 and 8 $\mu$m, and the best-fit SEDs from the two SED template
sets.  Filled symbols are detections and the error bars are smaller than the symbols.
Open symbols with arrows are 1 $\sigma$ upper limits.  The open symbol with
an error bar is the 0.65 $\sigma$ $K_s$-band measurement.  Note the vertical 
scale of the top panel.  The bottom panel 
shows the distributions of the $\chi_\nu^2$, $A_V$, and probability of the fits.  
\label{nir_photoz}}  
\end{figure}

Figure~\ref{nir_photoz} shows the results of our photometric redshift analyses.
Both SED template sets give best fits at redshifts between 6 and 7.  
The BC03 best-fit is a galaxy at $z=6.9$ that formed in a single burst of star formation
and has now aged 0.7 Gy with an extinction of $A_V=1.8$.  The empirical SED set 
provides a best fit at $z=6.4$, with an elliptical type and $A_V=0.5$.
The BC03 set systematically gives higher extinctions at all redshifts,
likely because of the limit in the galaxy age, i.e., an unreddened galaxy
that is sufficiently red to fit the observed SED may be too old for the age of
the universe.  Nevertheless, the $\chi_\nu^2$ distributions from the two
SED sets are consistent with each other.  In our subsequent analyses, we
adopt the BC03 result as it includes an age limit for the galaxy and provides 
physical quantities such as the age and stellar mass.  The 68\% confidence range (1 $\sigma$ 
for a Gaussian distribution) from the integration of the BC03 probability function
is $z=6.0$--7.4.  We note that redshifts between 4 and 5 are
not entirely ruled out, although the fit is poorer here.  On the other hand, 
$z<3$ is safely ruled out by the strong limits from the NICMOS and MOIRCS 
non-detections.

Comparing to the photometric redshift analyses in W07, the probability distribution 
now becomes narrower, which is an important improvement of this work.
On the other hand, the minimum $\chi_\nu^2$ increases, indicating that it is 
generally difficult to explain the red color between 1.6 and 3.6 $\mu$m.
We tested photometric redshifts at $z>10$ and obtained nearly perfect fits at $z>17$ 
(i.e., a Lyman break between $K_s$ and 3.6 $\mu$m).  However, this high redshift 
cannot explain the observed SED in the FIR and radio.  We do not consider 
$z>10$ in this paper.

\subsection{Millimetric Redshift}\label{sec_radio_ir_photoz}
We used the radio and FIR portion of the SED for another photometric
redshift estimate (the millimetric redshift estimate).  
Because there is not a clear correlation between the
radio/dust SEDs and the stellar SEDs of galaxies, this millimetric 
redshift estimate was carried out independently of the above NIR
photometric redshift.  The earliest version of such millimetric
redshift estimates was carried out using the spectral index between two wavebands 
in the radio and submillimeter \citep{carilli99, barger00, yun02} based on
the well known radio--FIR correlation in the local universe.
This method has larger errors caused by the uncertainty in the dust temperature. 
Advanced versions utilize full SED fitting in the radio and FIR
as the amount of available data increases.

Here we used all the data listed in Table~\ref{tab2} between
24 $\mu$m and 20 cm, including the two non-detections at 70 and 
160 $\mu$m.  These two non-detections play a key role in ruling out low redshifts of 
$z<3$ but do not provide strong constraints at $z>4$.
We used two sets of radio and FIR SED templates.
The first set contains model SEDs of Arp 220 (ultraluminous starburst
with cooler dust emission), NGC 6090 (luminous starburst),
M 82 (low luminosity starburst with warm dust) and M 100 
(normal spiral) adopted from \citet{silva98}, and Mrk 231 (ultraluminous dusty 
active galactic nucleus, AGN, with warm dust) derived from the photometry in 
NED\footnotemark[11].  The second SED set includes all of the 105 SED 
models in \citet[CE01]{chary01}.

\footnotetext[11]{The NASA/IPAC Extragalactic Database 
(NED) is operated by the Jet Propulsion Laboratory, California Institute of 
Technology, under contract with the National Aeronautics and Space Administration.}

\begin{figure}
\epsscale{1.1}
\plotone{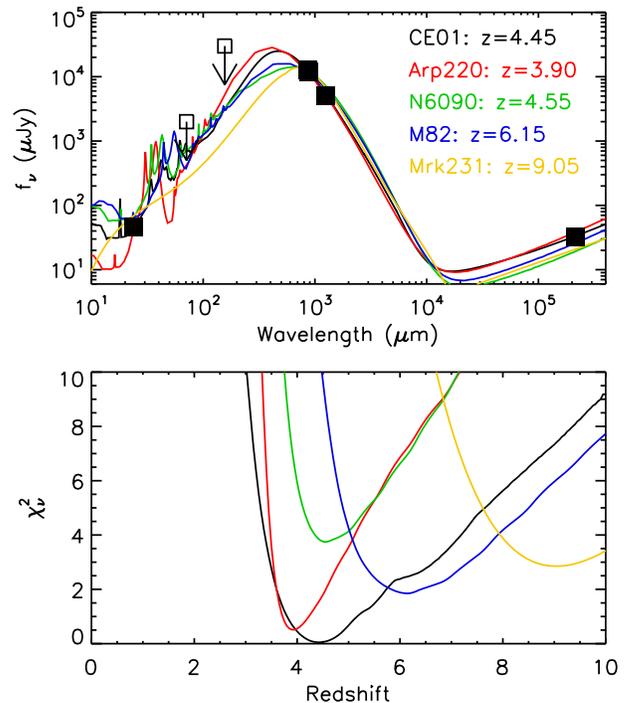}
\caption{Radio and FIR photometric redshift results.  The top panel shows the observed SED of
GOODS 850-5 and the best-fit SEDs.  The bottom panel shows the reduced
$\chi^2$ of the fits vs.\ redshift.  In the bottom panel, the black curve only shows the
best-fit among all 105 model SEDs in CE01 at each redshift.
\label{fig_radio_ir_photoz}}  
\end{figure}

In Figure~\ref{fig_radio_ir_photoz}, we present the best-fit SEDs and 
the redshift distribution of $\chi_\nu^2$ for each SED template.  A low redshift 
of $z<3$ is ruled out and this agrees with the NIR photometric redshift. 
The best-fit with the Silva et al.\ models
comes from Arp 220 at $z=3.9$.  This is perhaps not too surprising
since Arp 220 has been commonly considered as the local analogue to
high-redshift SMGs.  The next best fit comes from M 82.   It is at $z=6.2$ and
the fit is slightly poorer than that with Arp 220.  The fit with the spiral template M 100 has 
$\chi_\nu^2>10$ everywhere and is not shown in the figure.
The best-fit AGN template has a redshift of 9.1, which was discussed in
W07.  The best-fit of the CE01 templates is at $z=4.5$ and has a
$\chi_\nu^2$ ($\sim0$) lower than that from the Silva et al.\ models.  This unreasonably
low $\chi_\nu^2$ is perhaps a result of the very wide range of models (105 of them).
Generally speaking, the CE01 models provide good fits in the
$z\sim3.5$--5.5 range.  D08 used both the two-waveband spectral index method and 
the SED fitting method (also with the CE01 templates, but only with
the 24 $\mu$m, 850 $\mu$m, 1.25 mm, and 20 cm bands) and obtained $z\sim3.7$
and $z\sim3.3$ respectively.  Our photometric redshift results are slightly higher 
than those in D07, likely due to the introduction of the two upper limits at 
70 and 160 $\mu$m, which strongly disfavor $z<3.5$.  Nevertheless, the results here and 
those in D08 are still broadly consistent.

\subsection{$z\sim4$ or $z\gtrsim6$?}\label{4_or_6}

We are now faced with two different photometric redshift results.  
The NIR photometric redshift suggests $z\sim6$--7, while the millimetric
redshift suggests $z\sim4$.  We favor the NIR redshift of $z>6$ because we
believe that our knowledge about the stellar photospheric emission of galaxies is more 
robust than that about the dust and radio emission.  This is not only because the stellar 
spectral synthesis models (such as the Bruzual \& Charlot models) and the empirical 
spectra (such as the Coleman, Wu, \& Weedman templates) have been widely adopted
and tested in numerous extragalactic studies from $z\sim0$ to $z>7$,
but also because of the various uncertainties in the radio/IR photometric redshift.
Below we discuss each of the uncertainties.
 
First, the two-waveband spectral index method suffers from the uncertainty
in dust temperature.  Second, the stellar contamination in the 24 $\mu$m band
is quite uncertain at $z>4$.  All NIR photometric redshift models for GOODS 850-5
at $z>4$ predict observed 24 $\mu$m stellar photospheric fluxes that are much weaker
than those in the Silva et al.\ and CE01 models.  If we decrease the stellar emission in
the Silva et al.\ models to match the observations, all fits become better than those in 
Figure~\ref{fig_radio_ir_photoz}.  In particular, the M 82 fit at $z\sim6$ slightly shifts to
$z\sim5.5$ and becomes better than the Arp 220 fit at $z\sim4$.  
Third, and most importantly, there is a fundamental problem in the millimetric redshift, 
which is the uncertainty in the radio--FIR correlation.  

The radio--FIR correlation is a tight correlation observed for local
galaxies \citep{condon92}, but the detailed physical mechanism of this correlation
is still unclear.  Several groups have attempted to measure this correlation
on different samples of high-redshift galaxies with various methods, 
and the results remain controversial \citep{appleton04,boyle07,vlahakis07}.
If high-redshift SMGs are brighter in the radio, as suggested by Vlahakis et al.,
then the redshift of GOODS 850-5 should be higher than that inferred from the local
radio--FIR correlation.  In addition to the uncertainties in observational 
determinations of the correlation, there are also physical reasons 
to suspect that the radio--FIR correlation at high redshift would be different than 
that in the local universe.  For example, the higher energy density of the cosmic 
microwave background radiation at high redshift leads to a stronger inverse-Compton 
cooling for the cosmic-ray electrons \citep{condon92},
which suppresses the radio emission.  If this is the dominant effect at high redshift,
the correct redshift of GOODS 850-5 should be lower than that inferred from the
local correlation.  On the other hand, galaxies tend to have smaller sizes at high redshift.  
This may enhance the interstellar magnetic field and the synchrotron radiation, 
but the total radio emission can be further complicated by ionization and bremsstrahlung
loss \citep{thompson06}.

\begin{figure}[h!]
\epsscale{1.1}
\plotone{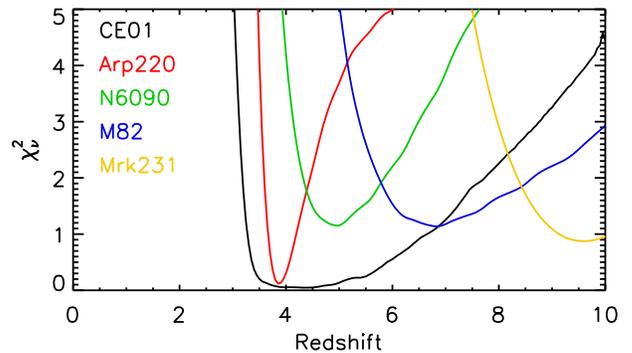}
\caption{FIR-only photometric redshift results. Description the curves is as in
Figure~\ref{fig_radio_ir_photoz}.  The plot legend (except for CE01)
are ordered by dust temperature: Arp 220 has the coolest dust and Mrk 231 has the
warmest dust.  Note that the vertical scale is different than that in Figure~\ref{fig_radio_ir_photoz}.
This plot shows that the observed FIR SED of GOODS 850-5 can be reasonably
fitted with various templates from $z\sim4$ to 10.
\label{fig_ir_photoz}}  
\end{figure}

Unfortunately, at least in the case of GOODS 850-5, the result of the 
millimetric redshift highly relies on the validity of the radio--FIR
correlation.  To show this, we repeated the same photometric redshift fitting
but did not use the 20 cm data point.  The resultant $\chi_\nu^2$ distributions are shown in 
Figure~\ref{fig_ir_photoz}.  In this experiment, all the fits have $\chi_\nu^2$ 
around 1.0 (the expectation value) between $z\sim3$ and $z>10$.  
This shows that the redshift of GOODS 850-5 highly depends on its dust temperature
when the radio--FIR correlation is not assumed.  
While it is established that many high-redshift SMGs have cool dust SEDs
similar to that of Arp 220, a small number of SMGs have warmer, M 82-like SEDs 
(e.g., \citealt{clements08}, also see Figure~2 in W07).
Because of this, we cannot rule out higher redshifts of $z>6$ and an M 82-like
SED for GOODS 850-5.

\subsection{Summary on the Redshift}

A low redshift of $z<3$ is ruled out by both the NIR and millimetric
redshifts for GOODS 850-5.  The NIR photometric redshift suggests
$z\sim6$--7, but the millimetric redshift suggests $z\sim4$.
The later strongly relies on the validity of the local radio--FIR correlation.
If we do not assume the local radio--FIR correlation, redshifts between 3 and 10 
seem equally possible for the observed FIR SED of GOODS 850-5.  
We consider the $z\gtrsim6$ NIR photometric redshift as a more likely one,
but $z\sim4$ is still a possibility for GOODS 850-5.

\section{Properties of GOODS 850-5}\label{sec_property}

\subsection{Active Galactic Nucleus?}
The possibility of the existence of an AGN was discussed by W07.  
The lack of an X-ray counterpart in the 2 Ms \emph{Chandra} image \citep{alexander03} 
rules out a Compton-thin AGN at $z<3$.  An AGN with an X-ray luminosity  
$\sim10^{42}$ erg s$^{-1}$ at $z>3$ cannot be detected by \emph{Chandra} and 
therefore is still possible.  If the radio and FIR emission of GOODS 850-5
is entirely powered by a Compton-thick dusty AGN like Mrk 231, its redshift would be $z\sim9$
(\S~\ref{sec_radio_ir_photoz}, W07), which is less likely.  Furthermore, our new observation
in the NIR shows a strong spectral break around 2--3 $\mu$m, which is
an important feature of old stellar populations.  On the other hand, QSOs typically
have power-law SEDs across a wide wavelength range from the UV to the NIR and
do not show strong spectral breaks like this.  We conclude that there is no evidence 
for an AGN in any of the observations.

\subsection{Stellar Population}
The use of the BC03 models and the latest Hyperz allows us to estimate the stellar 
masses and ages.  The SED of GOODS 850-5
is nearly a power law in the IRAC bands but shows a clear spectral break between
1.6 and 3.6 $\mu$m.  This can hardly be explained by a normal extinction curve.
A spectral break in the unreddened stellar SED is required, and the most natural
spectral break is the 4000 \AA\, Balmer break, which is a signature of older stars.
Hyperz agreed with this interpretation and all the best-fit models between $z=4$ and 10 have 
nearly the age of the universe.  For example, at the best fit of $z=6.9$,  the age of the universe is 
0.79 Gyr and the best-fit model has an age of 0.7 Gyr .  The minimum age 
with good fits ($\chi_\nu^2$ probability $>0.5\times$ the best-fit one) is 0.5 Gyr,
corresponding to a formation redshift of 10--14.  Because there are no detections in
the $K_s$ and F160W bands, the actual strength of the Balmer break in GOODS 850-5 is
unknown.  Therefore we do not think we can constrain its age with sufficient accuracy.
Nevertheless, we can conclude that it requires a well established stellar population 
with an age that is significantly large compared to the cosmic time to explain
the observed SED at $<10$ $\mu$m.  

\begin{figure}
\epsscale{1.1}
\plotone{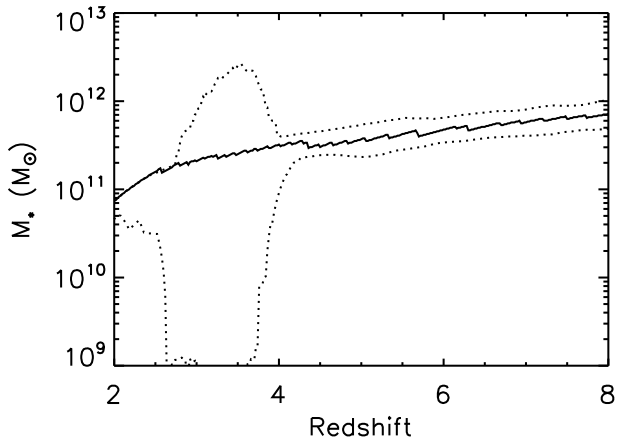}
\caption{Stellar mass vs.\ redshift, derived from Hyperz with the
BC03 models.  Solid curve is the best-fit result.  Dotted curves 
are the maximum and minimum stellar masses when the 
$\chi_\nu^2$ is larger than the minimum $\chi_\nu^2$ by 1.0
(see Figure~\ref{nir_photoz} for the distribution of minimum $\chi_\nu^2$),
showing how the stellar mass changes when the fit is perturbed
around the best-fit.  At $z<4$, the fits become very poor although the best-fit
mass still follows the trend at $z>4$.  At $z\sim6.9$, the minimum and 
maximum range is approximately $\pm40\%$ around the best fit.
There is another factor of 2 uncertainty caused by the uncertainty
in the stellar population model (see text).
\label{fig_mass}}  
\end{figure}

Since the observed IRAC fluxes are likely dominated by old stars with little
AGN contributions, they can be used to make a measurement of the stellar masses.
Figure~\ref{fig_mass} shows the Hyperz/BC03 best-fit masses as a function of
redshift, and the uncertainty range.  At $z=4$ and 6.0, the best-fit masses are
3 and $5\times10^{11}$ $M_{\sun}$, respectively.  The range allowed by the
photometric redshift fitting is quite narrow at $z>4$.   We therefore believe that the true 
uncertainty in the mass estimate is more likely from the stellar population
model.  In BC03, we can see that the mass-to-light ($M/L$) ratio is a strong function of 
initial mass function (IMF) and a weaker function of metallicity.  The adopted IMF in 
BC03 is the \citet{chabrier03} IMF, which is more top-heavy than the standard 
\citet{salpeter55} IMF.  If the IMF is Salpeter, the stellar mass of GOODS 850-5 would 
be approximately 2 times larger than the above values.  Another important factor
is the likely existence of the thermally pulsing asymptotic giant branch 
\citep[TP-AGB,][]{maraston05} phase, which is not included in the BC03 models.  
Such a phase decreases the $M/L$ ratio for a young stellar population in the rest-frame 
NIR. \citet{maraston06} found that the inclusion of the TP-AGB stars decreases the
mass by $\sim60\%$ on average, compared to the BC03 models.  
This would decrease the stellar mass of GOODS850-5 to $3\times10^{11}$ $M_{\sun}$
at $z\sim6$, but this is still a very large mass.

\subsection{IR Luminosity and Star Formation Rate}
\begin{figure}
\epsscale{1.1}
\plotone{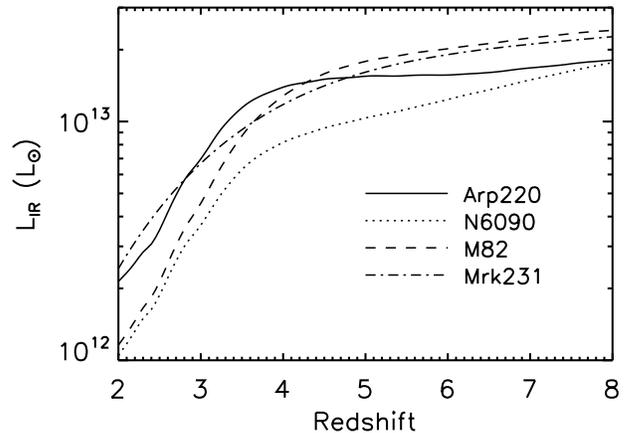}
\caption{Total IR luminosity for the best-fit models vs.\ redshift.  The IR luminosity is
derived from SED fitting to data between 24 $\mu$m and 1.25 mm.  The result
from the M 100 SED is not shown since M 100 does not fit the observations
($\chi_\nu^2 > 10$ for all redshifts).  
\label{fig_lir}}  
\end{figure}

The data at $>24$ $\mu$m allow us to robustly determine the total IR luminosity
(integrated from 8--1000 $\mu$m, e.g., \citealt{sanders96}). 
Figure~\ref{fig_lir} shows the IR luminosity of the best-fit \citet{silva98} models
(FIR only, \S~\ref{4_or_6} and Figure~\ref{fig_ir_photoz}) as functions of redshift and 
SED type.  We consider two redshifts here: $z\sim4.0$, as suggested by the millimetric
redshift estimate, and $z\sim6.9$, as suggested by the NIR photometric redshift.
At $z\sim4$ the best-fit model is Arp 220 and the best-fit IR luminosity is 
$1.4\times10^{13}$ $L_{\sun}$.  At $z\sim6.9$ the best-fit model is M 82 
and the luminosity is $2.6\times10^{13}$ $L_{\sun}$.  These are all comparable to
the result directly inferred from the submillimeter flux using the standard formulas 
(e.g., \citealt{blain02},  $L_{\rm IR} = 1.9 \times 10^{12} S_{850 \mu \rm m} L_{\sun}/$mJy,
which is $2.5\times10^{13}$ $L_{\sun}$ for GOODS 850-5).

The star formation rate for GOODS 850-5 can be estimated with 
the relation $\dot{M}=1.7\times10^{-10} L_{\rm IR}/L_{\sun}$ \citep[e.g.,][]{kennicutt98}.
The results are 2400 $M_{\sun}$ yr$^{-1}$ for $z\sim4$ and 4400 $M_{\sun}$ yr$^{-1}$
for $z\sim6.9$.  This assumes the standard Salpeter IMF.

\begin{figure}
\epsscale{1.1}
\plotone{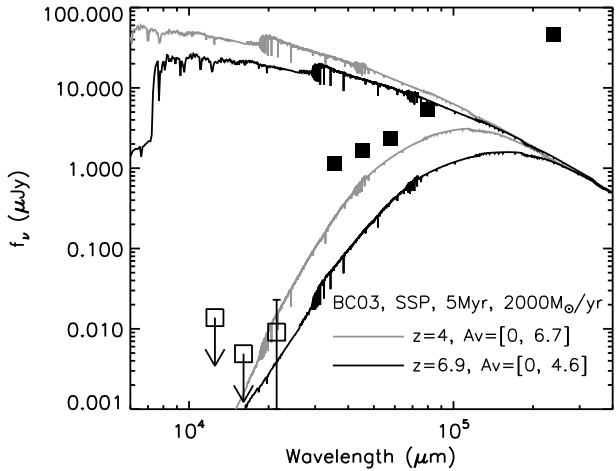}
\caption{NIR SEDs of ongoing starbursts suggested by the IR luminosity of GOODS 850-5
at $z=4.0$ and 6.9.  Two curves are shown for each redshift: one with no extinction, 
derived from a BC03 instantaneous burst model with an age of 5 Myr and a star formation
rate of 2000 $M_{\sun}$ yr$^{-1}$; the other with an extinction that is required to minimally 
hide the burst in the F160W and $K_s$ bands such that the hidden burst contributes 
$<50\%$ to the observed 1 $\sigma$ upper limits.  The squares are the observed SED of 
GOODS 850-5 at $>1$ $\mu$m.  \label{fig_hidden_burst}}  
\end{figure}

It is interesting to compare the above star formation rates with the non-detections in the 
NIR.  Using the UV luminosity versus star formation rate conversion in \citet{kennicutt98}, 
$\dot{M}=1.4\times10^{-28}L_{\rm UV}/$(\rm erg s$^{-1}$ Hz$^{-1}$),
and a minimum IR star formation rate of 2000 $M_{\sun}$ yr$^{-1}$, we found an
unattenuated UV (1500--2800 \AA) luminosity of $1.4\times10^{31}$ erg s$^{-1}$ Hz$^{-1}$.  
At $z=4$--6.9, this corresponds to approximately 40--20 $\mu$Jy at 1.6 to 2.1 $\mu$m, 
assuming a young burst SED from BC03.  The observed F160W and $K_s$ flux upper limits then imply 
the extinctions for the starburst component to be at least $A_V\sim6.7$--4.6.  
Figure~\ref{fig_hidden_burst} shows the SEDs of such hidden starbursts at $z=4$ and 6.9.
The lower limits for the extinctions of the starburst component are significantly higher 
than those derived from the optical/NIR SED.     
We will come back to this in \S~\ref{sec_coexist}.

\subsection{Nature of the 24 $\mu$m Emission}
All the above discussions on the existence of an AGN, 
the stellar mass, and the IR luminosity depend (to various degrees)
on whether the observed 24 $\mu$m emission is dominated by
emission from stellar photosphere or from dust (see the extensive discussion on the 24 
$\mu$m emission from HUDF-JD2 in \citealp{mobasher05}).  
Roughly speaking, the observed 24 $\mu$m flux of GOODS 850-5 is 
unlikely to be dominated by stellar emission, as this would
require $A_V>8$ at $z\sim6$ to explain the observed red 24-to-8.0 $\mu$m
color without dust emission.  If we include the 24 $\mu$m band in the
optical/NIR photometric redshift analyses, the fit becomes poor
at all redshifts.

We can quantify the strength of stellar radiation
at 24 $\mu$m using photometric redshift fits 
without including the 24 $\mu$m data.  The best-fit BC03 models
in \S~\ref{sec_nir_photoz} imply 24 $\mu$m stellar emission of 
4.6 and 2.9 $\mu$Jy at $z\sim4$ and 6.9, respectively, corresponding
to 10\% and 6\% of the observed 24 $\mu$m flux.  These seem 
unusually small, especially at $z\sim6.9$, where 24 $\mu$m 
corresponds to rest-frame 3 $\mu$m.  We can compare these with
local templates in \citet{silva98} to see if they make any sense.
Table~\ref{tab3} compares the ratios of $L_{\rm IR}/M_{\star}$ and
the ratios of dust-to-stellar radiation at 24 $\mu$m for Arp 220, NGC 6090, 
M 82, and GOODS 850-5, at $z\sim4$ and $z\sim6.9$. 
We found that once the dust-to-star ratios in the observed 24 $\mu$m emission
is normalized by the $L_{\rm IR}/M_{\star}$ ratios, GOODS 850-5
is not different from local galaxies.  It is also interesting to note that
at $z\sim4$ and 6.9 the normalized dust-to-star ratios are similar to
Arp 220 and M 82, respectively.  This is consistent with what we found
in the photometric redshift analyses:  GOODS 850-5 is similar to Arp 220
if it is at $z\sim4$ but more similar to M 82 if $z>6$.  

\begin{deluxetable}{llccc}[h!]
\tablecaption{Dust and Stellar Contributions at 24 $\mu$m\label{tab3}}
\tablehead{\colhead{$z$}  & \colhead{Galaxy} 
& \colhead{$L_{\rm IR}/M_{\star}$} & 
\colhead{$S_{24}^{\rm dust}/S_{24}^{\star}$} &
\colhead{$(S_{24}^{\rm dust}/S_{24}^{\star})^{\prime}$} }
\startdata
4 \\
\hline
& Arp 220			&	14	&  4.8  	& 0.34\\
& NGC 6090 		& 	1.5	&  1.6  	& 1.1 \\
& M 82 			& 	0.2	&  2.7  	& 13 \\
& GOODS 850-5	&	46	&  15  	& 0.33\\ 
\hline
6.9 \\
\hline
& Arp 220			&	14	&  0.08  	& 0.0057\\
& NGC 6090 		& 	1.5	&  0.07  	& 0.047\\
& M 82 			& 	0.2	&  0.08  	& 0.40\\
& GOODS 850-5	&	52	&  9.1  	&0.18
\enddata
\tablecomments{$L_{\rm IR}/M_{\star}$ is in solar unit.  
$(S_{24}^{\rm dust}/S_{24}^{\star})^{\prime} = 
(S_{24}^{\rm dust}/S_{24}^{\star}) / (L_{\rm IR}/M_{\star})$.}
\end{deluxetable}

There are two subtle issues in the above comparison.  First, the 
underlying assumption for the normalization is that the $M/L$  ratios
are similar for the galaxies at 24 $\mu$m (rest-frame 3.0--4.8 $\mu$m
for the redshifts of interest).  Given the long wavelengths, this is
acceptable.  Second, the estimates of the IR luminosity for GOODS 850-5
slightly depends on how much of the observed 24 $\mu$m emission 
comes from the warmer dust components.  However, from 
Figure~\ref{fig_lir} we can see that the uncertainty is well within
a factor of 2 and the IR luminosity is mostly determined by the 
three measurements in the millimeter and submillimeter.
Therefore the 24 $\mu$m measurement can still be considered
as a semi-independent check for all the analyses in longer and
shorter wavelengths.  The fact that the extrapolations of the optical/NIR 
photometric redshift analyses agree well with the millimetric one at
24 $\mu$m (in terms of dust-to-star ratio) does show an excellent self 
consistency.  It is fair to conclude that the unusually large dust-to-star 
ratio in the observed 24 $\mu$m emission of GOODS 850-5 is simply 
a consequence of its extremely large IR luminosity.  

Two question arises once we establish that $\gtrsim90\%$ of the 
observed 24 $\mu$m emission comes from dust:  (1) Does this affect
our millimetric redshift analyses?  (2) Does the dust
emission extend to the IRAC bands and bias our stellar mass
estimates?  To examine (1), we decreased the stellar contribution
in the Silva et al.\ templates by factors of 2 to a few tens and repeated the 
photometric redshift fitting.  We found that the fits improve slightly.
The best-fit redshift with Arp 220 does not change significantly and
the best-fit redshift with M 82 decreases from $\sim6$--7 to $\sim5.5$--6.
Thus, the general conclusions on the millimetric redshift
and the IR luminosity are both fairly insensitive to the assumed stellar 
contribution to the 24 $\mu$m flux.  

The answer to (2) is negative as well.  The dust spectral slope for 
starbursts is extremely steep at $<1.6$ $\mu$m (IRAC bands for $z>4$),
and the blue end of the dust emission does not contaminate the IRAC fluxes.  
This is still true even if the observed 24 $\mu$m emission comes from a 
dusty AGN with dust much warmer than starbursts.  \citet{mobasher05} 
compared the observed 24 $\mu$m emission from HUDF-JD2 with dusty AGN 
templates Mrk 231 and NGC 1068.  We follow the procedure in Mobasher et al.\
and compare the SEDs of GOODS 850-5 and NGC 1068.  We found that
the dust emission only contributes to the IRAC fluxes at $z<4$.  
At $z>6$, similar to what Mobasher et al.\ concluded for HUDF-JD2 at $z=6.4$, 
the dusty AGN does not bias the IRAC stellar mass estimates.
We do not consider Mrk 231 here as its SED in the literature is contaminated by 
stellar emission at rest-frame $<3$ $\mu$m (as pointed out by Mobasher et al.) 
and hence it does not provide a definite answer.  Moreover, as shown in 
our photometric redshift analyses, to explain the large millimeter fluxes of GOODS 850-5
with a Mrk 231-like warm dust, it would require unlikely high redshifts of 
$z\sim8$--10.  In short, we conclude that the dust emission at 24 $\mu$m
is unlikely to bias our IRAC-based stellar mass estimates regardless of the existence
of a dusty AGN.

\subsection{Coexistence of Old Stars and a Starburst}\label{sec_coexist}
The properties of GOODS 850-5 are puzzling.
There appear to be two inconsistencies that are related to each other.
(1) While the IR luminosity implies an intensive ongoing starburst and the formation
of a young galaxy, the NIR SED suggests that most of the observed stellar 
radiation comes from old stars without any detectable rest-frame UV radiation. 
(2) Our interpretation of the NIR SED suggests $A_V\lesssim2$, but $A_V>4.6$ is 
required to hide all the young stars.

One conceivable way to solve this is to see if $A_V>4.6$ with young stellar populations
could provide reasonably good fits in the photometric redshift.  If yes, we may argue 
that the entire young galaxy is behind an $A_V>4.6$ dust screen 
and that the observed IRAC fluxes indeed mostly come from young 
stars.  To investigate this possibility, we limited the 
photometric redshift fitting to the BC03 models with $A_V=4.6$--10, $z>4$, and ongoing 
star formation.  At $z\sim4$, fits with such highly extinguished stellar populations 
all have $\chi_\nu^2>2.7$, which is significantly poorer than the fits with $A_V\lesssim2$.
At $z>5$, the $\chi_\nu^2$ becomes larger than 5 and the models fail to fit the data.
We conclude that $A_V>4.6$ for a young galaxy is a less likely case, although not fully 
ruled out for $z\sim4$.  It is more likely that the observed IRAC fluxes mostly come from
a relatively old stellar population with a moderate extinction of $A_V\lesssim2$.

Another possible and indeed sensible scenario to explain the above inconsistencies 
is that the star forming region (with $A_V>4.6$) in GOODS 850-5 is different from the 
region that produces most of the observed IRAC fluxes ($A_V<2$).  
GOODS 850-5 is not resolved by the SMA ($\sim2\farcs2$ beam FWHM, S/N=8.6) 
and by the VLA ($\sim1\farcs7$, S/N=8.2).  These imply an uncertainty of 
$\sim0\farcs2$ (beam FWHM divided by S/N) for its beam-convolved size
and $\sim0\farcs8$ for the upper limit of its intrinsic size.
The upper limit corresponds to $\sim4.5$ kpc at $z=6.9$ and is consistent 
with the CO sizes of low-redshift SMGs \citep{tacconi06}. 
On the other hand, the IRAC fluxes suggest a massive stellar population of 
$M_{\star} \sim10^{11.5} M_{\sun}$, which may be spatially offset from the
starburst region and may have less extinction.

An interesting object for comparison is the prototypical dusty starburst ERO, 
Hu-Ridgway 10.
\citet{stern06} found that its 10 $\mu$m silicate feature implies $A_V\sim11$
but its optical/NIR SED implies a much smaller $A_V\sim2.4$.  
They ascribed this to a heavily obscured starburst in Hu-Ridgway 10, similar to our hypothesis
for GOODS 850-5.  Unfortunately, the current data of GOODS 850-5 do not allow 
us to use the silicate feature for an extinction measurement.
Another possibility to test the above two-component hypothesis for GOODS 850-5 is to see if 
we can fit the observed NIR SED with two components.  We found that the photometric
redshift fitting becomes poorer at all redshifts if we include hidden bursts shown
in Figure~\ref{fig_hidden_burst}.  The increase in $\chi_\nu^2$ is from $\sim0.2$
at $z=4$ to $\sim0.5$ at $z=6.9$.  However, we note that the actual extinction for
the starburst component can be higher than the minimally hidden burst shown in 
Figure~\ref{fig_hidden_burst}.  Once we increase the extinction for the burst component
by $\sim2$--3 (still less than the extinction in Hu-Ridgway 10), the photometric redshifts 
reduce to the ones shown in \S~\ref{sec_nir_photoz} at all redshifts.  
In other words, the current data are insufficient for further testing the two-component
hypothesis.

We conclude that a massive galaxy with an old stellar population but with 
some dust screening hosting a much more dusty nuclear starburst can explain 
the observations of GOODS 850-5.

\section{Discussion}\label{sec_discussion}

For the stellar masses and star formation rates
estimated at $z\sim4$ and 6.9, the mass build-up times are both
$\sim110$ Myr, respectively corresponding to 7\% and 14\% of their cosmic times.
It is clearly more difficult to produce the stellar mass for the $z\sim6.9$ case,
since there is less time available for the formation of the galaxy.
Given that all the stars we see are old, it requires an intensive but short burst of
$>10^3 M_{\sun}$ yr$^{-1}$ at $z>10$.  This also suggests that GOODS 850-5
has undergone two distinct bursts of star formation: one 
produced the old stars observed in the IRAC bands, and the other is producing the 
observed dust emission in the MIPS and (sub)millimeter bands.

The estimated mass of $M_{\star}\sim10^{11.5}M_{\sun}$ is similar to typical SMGs 
at $z<4$ \citep[e.g.,][]{smail04,dye08} but is more massive than other galaxies
observed at $z>6$.  The most massive Lyman-break (UV luminous) selected galaxies at 
$z\sim6$--7 have stellar masses of $\sim10^{10}$--$10^{10.5}$ $M_{\sun}$ 
\citep{labbe06,yan06,eyles07}.
GOODS 850-5 is 10 times more massive than these optically selected galaxies, 
suggesting that the very massive galaxies at these redshifts may be still in dusty
starbursting phases and therefore may be missed by deep optical surveys.

We can estimate the limit for the number of massive, dust-hidden galaxies at $z>6$.
The photometric redshift uncertainty range for GOODS 850-5 is $z\sim6.0$--7.4
and our GOODS-N SCUBA survey area for sources brighter than GOODS 850-5 is
$\sim100$ arcmin$^2$.  Therefore the comoving volume of our survey for
sources similar to GOODS 850-5 is $\sim3\times10^5$ Mpc$^3$.  This implies
a number density of $\sim3\times10^{-6}$ Mpc$^{-3}$ for
massive objects of $M_{\star}\sim10^{11.5}$ $M_{\sun}$.  It is unclear how a
radio-faint submillimeter selection is related to stellar masses.  However, the above 
density is probably a lower limit, since GOODS 850-5
is the first radio-faint SMG studied in our SMA survey and the survey is
not yet complete.  This lower limit is slightly larger than the maximum 
value suggested by \citet[see their Figure~11]{yan06} for optically selected 
galaxies at $z\sim6$, and is significantly larger than the values from the $\Lambda$CDM
hydrodynamic simulations of \citet{nagamine04} and \citet{night06}.
This emphasizes the importance of dusty galaxies at high redshift, but we 
clearly need a much larger high-redshift SMG sample to reach cosmologically 
meaningful conclusions.

It is interesting to ask whether current $\Lambda$CDM models can produce
$10^{11.5}$ $M_{\sun}$ galaxies at all at $z>6$.  In standard $\Lambda$CDM
models, massive galaxies form in later cosmic times (aka.\ ``hierarchical'' galaxy formation).  
On the other hand, evidence of ``cosmic downsizing''
\citep{cowie96} has been observed at many different wavelengths over
a broad range of redshifts, and this has been considered by some as 
anti-hierarchical.  The existence of ultraluminous SMGs at $z>2$ once put a strain on the 
$\Lambda$CDM models \citep{baugh05}.  Moreover, the stacking detection of a 
population of faint SMGs at low redshifts of $z\sim1$, which dominates the 
submillimeter EBL, further demonstrated the downsizing behavior in the SMG 
population from $z\sim4$ to $z=0$ \citep{wang06,serjeant08}.
Recent semi-analytic and hydrodynamic models are able to reproduce 
galaxies with $M_{\star}>10^{10}$ $M_{\sun}$ up to $z\sim5$--6
under the $\Lambda$CDM framework \citep{bower06,night06}.

At slightly higher redshifts of $z=6$--7, the $\Lambda$CDM N-body simulations of 
\citet{lukic07} provide a halo density of $\sim10^{-5}$ (Mpc/h)$^{-3}$ for the 
mass range of $10^{12}$--$10^{13}M_{\sun}$.  The density rapidly drops to
$10^{-9}$--$10^{-11}$ (Mpc/h)$^{-3}$ for the mass range of $10^{13}$--$10^{14}M_{\sun}$.
If we assume a matter-to-star mass ratio of $\gtrsim10$, the density of 
$\sim3\times10^{-6}$ Mpc$^{-3}$ that we derived for high-redshift SMGs seems to 
match the upper bound in the simulations.  This suggests that it is plausible
to find approximately one $z>6$ dark halo in our survey area that is massive enough to host a 
hyperluminous SMG.  However, whether the simulations can reproduce the observed 
intensive starburst and the large stellar mass within such a halo in a short cosmic time 
remains to be seen.

Lastly, can objects like GOODS 850-5 play a role in the reionization of the universe?
The latest 5 yr \emph{WMAP} result \citep{dunkley08} implies a reionization 
redshift of $11.3\pm1.4$ (for instantaneous reionization) and the massive 
old stellar population in GOODS 850-5 implies intensive star formation at 
$z>10$.  We use the standard formulation in \citet{madau99}, 
$\dot{\rho} = 0.013 \times f_{\rm esc}^{-1} \times [(1+z)/6]^3$ $M_{\sun}$ yr$^{-1}$ Mpc$^{-3}$,
where $\dot{\rho}$ is the minimum star formation rate density required 
for reionization, $f_{\rm esc}$ is the escaping fraction of ionizing photons, and a
Salpeter IMF is assumed.  With the number density estimated above, 
$f_{\rm esc}\sim0.1$, and assuming that the initial burst in GOODS 850-5 is as intensive 
as the current one, we found that the formation of GOODS 850-5 at $z>10$ 
contributed $<10^{-2}$ to the ionizing photons that are required for reioinization.
The fraction would be even smaller if the initial burst of GOODS 850-5 were dusty
(i.e., a smaller $f_{\rm esc}$).  This is consistent with the current picture that the universe 
is ionized by a large amount of low luminosity objects.

\section{Summary and Final Remarks}\label{sec_summary}
Our new ultradeep NIR observations reveal many unusual properties of GOODS 850-5.
It is detected by the SMA, IRAM, and VLA from the submillimeter to centimeter
wavelengths, and by \emph{Spitzer} between 3.6 and 24 $\mu$m, all with high
significance.  On the other hand, it is not detected in the $J$, F160W, and $K_s$ bands even 
with the nano-Jansky sensitivities.  We analyzed the 
photometric redshifts of GOODS 850-5.  The NIR photometric redshift suggests
a high redshift of $z\sim6.9$ and rules out $z<3$.  The millimetric redshift
also rules out $z<3$ and suggests $z\sim4$ if we assume the local radio--FIR correlation.  
Without this assumption, $z=4$ to 10 is equally possible for the observed IR SED.  
We conclude that $z\gtrsim6$ is more likely for GOODS 850-5
but $z\sim4$ remains a possibility. 

To explain the observed NIR SED of GOODS 850-5 and the IR luminosity requires 
an established stellar population that formed at a large look-back time
coexisting with an intensive ongoing starburst that is completely invisible in the 
rest-frame UV.  The old stars observed in the IRAC bands have a large mass of 
$M_{\star}\sim10^{11.5}M_{\sun}$ 
and are $\sim10$ times more massive than optically selected massive galaxies at $z>6$.  
The current burst of star formation seems to be compact with a total IR luminosity 
of 1.4--$2.6\times10^{13}$ $L_{\sun}$ and a star formation rate of 2400--4400 
$M_{\sun}$ yr$^{-1}$.
It is deeply enshrouded by an $A_V>4.6$ dust screen so its UV radiation is 
not detected by NICMOS at 1.6 $\mu$m and by Subaru at 2.1 $\mu$m.  
This large extinction required for the starburst component also makes it
difficult to directly confirm its existence in the IRAC bands and the
two-population hypothesis remains to be tested.

The high redshift of GOODS 850-5, if confirmed, will have important implications for
galaxy formation and evolution, as discussed in this paper and in W07. 
Its extreme faintness at $<2$ $\mu$m prevents any optical and NIR spectroscopy 
with current space and ground-based instruments.  A precise measurement
of its redshift will most likely come from millimeter ``redshift machines'' on large
telescopes such as the 110 m Green Bank Telescope.  Spectroscopy in the 
MIR and FIR with the \emph{Herschel Space Observatory} is another possibility.  
In general, the observations of GOODS 850-5 remind us that there is a
class of unexpected objects that are not included in the traditional picture of 
radio identified SMGs.  The SMA is likely to reveal more examples in the near future. 
Studies of cosmologically large samples of such high-redshift SMGs, however, will 
require combinations of next-generation instruments, such
as the Expanded VLA, Atacama Large Millimeter/Submillimeter Array, 
\emph{James Webb Space Telescope}, and large bolometer arrays on 
single-dish millimeter telescopes.

\acknowledgments
We thank L.\ Silva and R.\ Chary for providing the SED templates, 
the Subaru and \emph{HST} staff for help with observations and data reduction,
G.\ Morrison for providing us with the latest VLA image, C.\ Carilli, F.\ Owen, 
and H.\ Hirashita for very helpful discussions, and the referee for comments that
greatly improved the manuscript.
This work is supported by \emph{HST} grant HST-GO-11191,
the NRAO Jansky Fellowship program (W.-H.W.), NSF grants
AST 0239425 and 0708793 (A.J.B.) and AST 0407374 and 0709356 (L.L.C.), the 
University of Wisconsin Research Committee 
with funds granted by the Wisconsin Alumni Research Foundation, 
and the David and Lucile Packard Foundation (A.J.B.).


\end{document}